# Modeling of electron energy spectra and mobilities in semi-metallic $Hg_{1-x}Cd_xTe$ quantum wells

E.O.Melezhik*, J.V.Gumenjuk-Sichevska, F.F.Sizov

V.E.Lashkaryov Institute for semiconductor physics of NAS of Ukraine

Nauki 41 av., 03028 Kyiv, Ukraine

*emelezhik@gmail.com

**Abstract**. Electron mobility, energy spectra and intrinsic carrier concentrations in the n-type $Hg_{0.32}Cd_{0.68}Te$ / $Hg_{1-x}Cd_xTe$ / $Hg_{0.32}Cd_{0.68}Te$ quantum well (QW) in semi-metallic state are numerically modeled. Energy spectra and wave functions were calculated in the framework of the 8-band k-p Hamiltonian. In our model, electron scattering on longitudinal optical phonons, charged impurities, and holes has been taken into account, and the mobility has been calculated by an iterative solution of the Boltzmann transport equation.

Our results show that the increase of the electron concentration in the well enhances the screening of the 2D electron gas, decreases the hole concentration, and can ultimately lead to a high electron mobility at liquid nitrogen temperatures. The increase of the electron concentration in the QW could be achieved *in situ* by delta-doping of barriers or by applying the top-gate potential. Our modeling has shown that for low molar composition *x* the concentration of holes in the well is high in a wide range of electron concentrations; in this case, the purity of samples does not significantly influence the electron mobility.

These results are important in the context of establishing optimal parameters for the fabrication of high-mobility$Hg_{1-x}Cd_xTe$ quantum wells able to operate at liquid nitrogen temperature and thus suitable for applications in terahertz detectors.

1. Introduction

The development of novel detectors operating in the terahertz (THz) spectral range is one of the important challenges of modern optoelectronics. THz detectors are characterized by such parameters as responsivity, operating speed, noise-equivalent power (NEP), spectral selectivity, and operation temperature. Different areas of application impose different requirements on THz detectors, hence there is a variety of detector types optimizing a specific subset of the above list of parameters.

In most practical applications, uncooled or moderately cooled (down to liquid nitrogen temperature) THz detectors are needed. One can distinguish between two kinds of such devices, namely, thermal ones, e.g., bolometers (in particular, hot-electron bolometers (HEBs)), and rectifying detectors, for instance, field-effect transistors (FETs) or Schottky barrier diodes (see, e.g., [1]). These devices can be suited to meet the requirements of a reasonably good sensitivity / speed, and they can have NEP comparable to other known uncooled detectors. Direct type detectors of that kind can be important in various civil areas, such as medicine, security, food control, etc.

Such devices as HEBs and FETs have a sensitive element, thin layer or channel, properties of which directly influence the performance and speed of the detector. Creating the channel as a quantum well (QW) is a good choice, because the momentum quantization in the QW growth direction allows for a significant reduction of the 2D electron heat capacity [2]. The requirement of a high sensitivity could be implemented by using a high-mobility material for the channel and cooling down to liquid nitrogen temperature, while low noise could be obtained by using a low-resistive channel. High operation speed could be realized for high-mobility channels, or for channels with a fast energy relaxation of the 2D electron gas (2DEG).

Mercury-cadmium-telluride (MCT) heterostructures are promising materials for creation of such channels. Depending on their parameters, the QWs can be characterized by high electron mobility and high electron concentrations even at liquid nitrogen temperatures. Depending on the molar composition $x$ and quantum well width $L$, a semi-metallic or semiconducting state can be realized in such QWs [3]. A semi-metallic state is characterized by the much higher conduction electron concentration at $T = 77$ K [4]. Thus, compared to the undoped semiconducting $Hg_{1-x}Cd_xTe$ QWs of the same width, semi-metallic QWs can have much lower resistivities and lower thermal noise. For that reason, we restrict our modeling to the case of semi-metallic $Hg_{1-x}Cd_xTe$ heterostructures.

To our knowledge, a systematic theoretical study of transport properties of semi-metallic $Hg_{1-x}Cd_xTe$ quantum wells at liquid nitrogen temperature is still lacking. The aim of the present work was to model the carrier energy spectra, the electron concentration and electron mobility in $Hg_{1-x}Cd_xTe$ QW at $T = 77$ K.

Numerical calculation of carrier energy spectra and wave-functions is carried out in the framework of the 8-band k-p model [4,5], which takes into account the strong band mixing and nonparabolicity of the dispersion law, and describes the transition between semi-metallic and semiconducting states driven by the change of the QW parameters. In the bulk $Hg_{1-x}Cd_xTe$ at the liquid nitrogen temperature, there are three efficient electron scattering mechanisms: inelastic scattering on longitudinal optical (LO) phonons, scattering on residual charged impurities, and electron-hole scattering (the latter two are elastic) [6]. To calculate the impact of these scattering mechanisms on the electron mobility in the quantum well, the linearized Boltzmann transport equation (LBTE) was solved iteratively. Solution of the LBTE allows one to take into account accurately the inelasticity of electron scattering and to describe how the carrier distribution function is perturbed by the applied electric field in the channel.

In our calculations, we have used the $Hg_{0.32}Cd_{0.68}Te/Hg_{1-x}Cd_xTe/Hg_{0.32}Cd_{0.68}Te$ quantum well as a model system. The composition $x = 0.68$ of the barrier layer was chosen since it is often used in real heterostructures [7, 8, 9] to minimize misfit strains between the well and barrier layers.

## 2. Energy spectra &intrinsic concentration

Calculations of energy spectra were performed in the framework of the 8-band k-p Hamiltonian [5] that is applicable to study both direct and inverted band orders in the quantum well. This model is described in detail in [4]. Here we give a brief outline of the dependencies of the energy spectrum, the Fermi level, and the intrinsic concentration on the QW width $L$ and composition $x$, obtained in the framework of the above model.

The central panel of Fig.1 presents the dependence of the QW energy spectrum on the well width $L$, while the left and right panels show the band dispersion for $L$=8 nm (semiconducting state) and $L$=20 nm (semi-metallic state), respectively. The dependence of the intrinsic concentration on $L$, calculated for the QW composition $x$=0.06, is shown in Fig.2. At this composition, the critical width (at which the band inversion occurs) is around $L$=12 nm (see Fig.1).

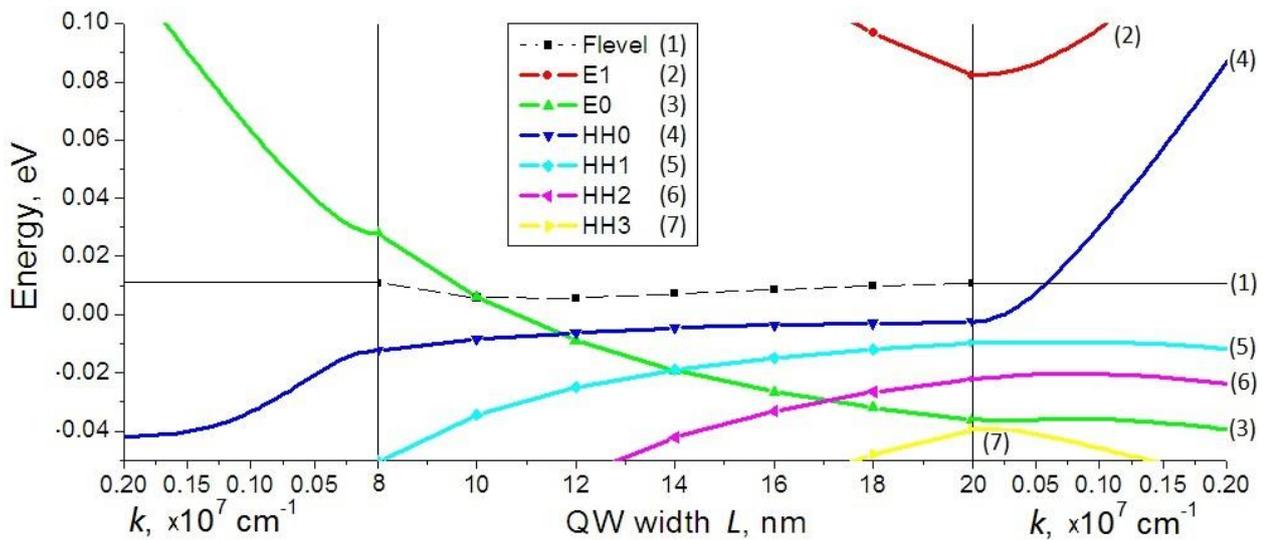

Figure 1. Dependence of the energy spectrum and the Fermi level in the QW with the composition $x$=0.06 on the well width $L$. Curves on the graph are marked as follows: the Fermi level – (1), E1 level – (2), E0 level – (3), HH0 level – (4), HH1 level – (5), HH2 level – (6), HH3 level – (7). Calculations are done for T = 77 K.

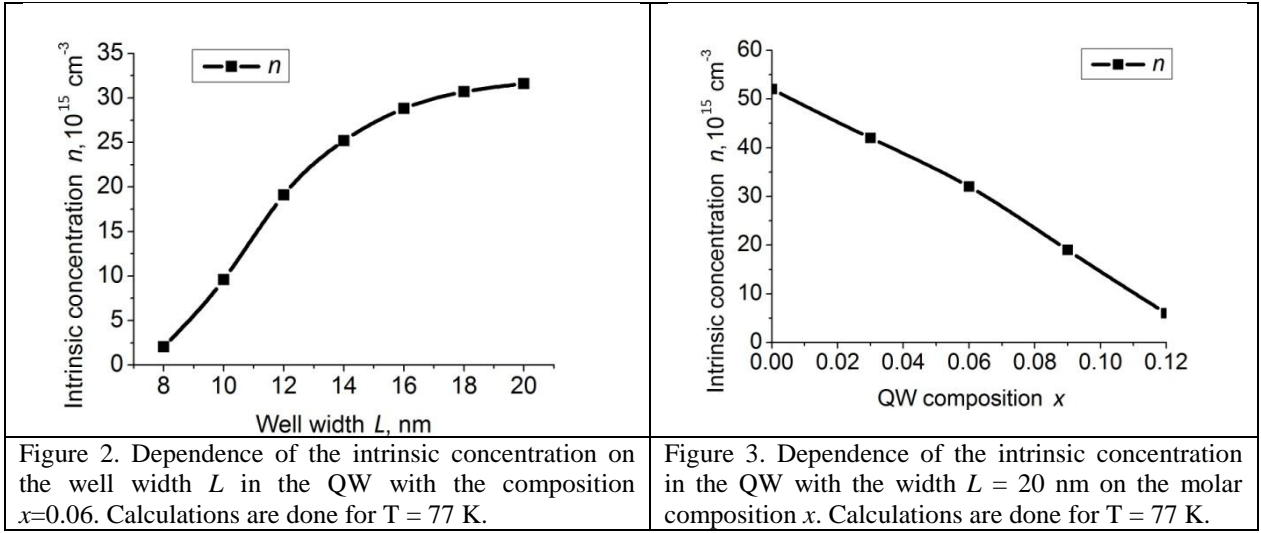

Figure 2. Dependence of the intrinsic concentration on the well width $L$ in the QW with the composition $x$=0.06. Calculations are done for T = 77 K.

Figure 3. Dependence of the intrinsic concentration in the QW with the width $L$ = 20 nm on the molar composition $x$. Calculations are done for T = 77 K.

From Fig.2 one can see that the electron concentration increases with the well width. This growth is explained by the transition of the system from the semiconductor to the semimetal state, when the well width is increased. Such transition opens the band gap for widths less than 12 nm. For larger well widths the system is semimetal, and the band-gap is absent, thus the intrinsic concentration exhibits a plateau at larger well widths.

Another important feature is that the local minimum of the Fermi level is situated near the critical width. One should note that for higher molar compositions like $x$=0.12, the critical width is around $L$~20 nm and the region of the concentration growth in Fig. 2 will be shifted to higher well widths.

The dependence of the electron concentration on the QW composition $x$ is presented in Fig.3. The corresponding energy spectrum is presented in Fig.4, where energy levels E0 and E1 have Γ6 symmetry, while HH0-HH3 and LH0 levels have Γ8 symmetry.

It should be noted that varying the composition can change the order of bands in the well. For $x$ = 0, the critical width for the HgTe QW is $L$=6.7 nm, while for higher concentrations it grows and exceeds $L$≈20 nm for $x$ = 0.12, which should be taken into account when fabricating MCT QWs for THz bolometer detectors. The carrier concentration in the well decreases with the increase of the composition $x$; this dependence is almost linear (Fig.3).

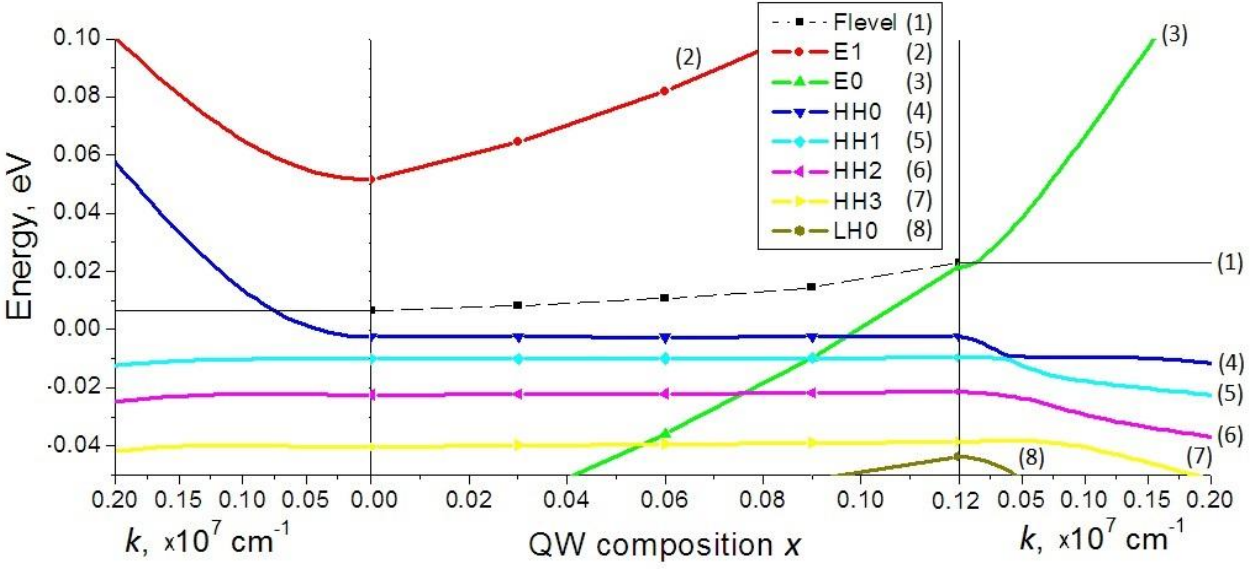

Figure 4. Dependence of the energy spectrum and the Fermi level of the QW with the width $L = 20$ nm on the composition $x$. Curves on the graph are marked as follows: the Fermi level – (1), E1 level – (2), E0 level – (3), HH0 level – (4), HH1 level – (5), HH2 level – (6), HH3 level – (7), LH0 level – (8). Calculations are done for T = 77 K.

## 3. Calculations of the perturbed distribution function

Our calculations of the distribution function are based on the solution of the Boltzmann transport equation. We followed the methodology of [10], which we adapted to the case of the nonparabolic dispersion law.

For the sake of simplicity, we studied the case when only the ground state energy level (mini-band) is populated and all scattering processes take place within this level. The energy of the bottom of the ground state level is denoted $E_0$.

### 3.1. The Boltzmann equation.

Consider the distribution function $f(r,k\ t)$ which gives the occupation probability of the state $|k\rangle$ by an electron in the volume element $dr$ at the position $r$ at time $t$. The rate of change of $f(r,k,t)$ with time is given by the familiar Boltzmann equation:

$$\frac{\partial f}{\partial t} = -\frac{1}{\hbar}\frac{\partial E}{\partial k}\frac{\partial f}{\partial r} - \frac{1}{\hbar}F\frac{\partial f}{\partial k} + I_c[f], \qquad (1)$$

where $E$ is the electron energy, $\boldsymbol{F} = -e\boldsymbol{E}_{ext}$ is the force acting on the electron due to the externally applied electric field $E_{ext}$, and $\hbar$ is the reduced Planck constant. The last term is the collision integral, which arises from the electron scattering and is given by:

$$I_c[f] = -\int \frac{d^2\boldsymbol{k}'}{(2\pi)^2}\{S(\boldsymbol{k},\boldsymbol{k}')f(\boldsymbol{r},\boldsymbol{k},t)[1-f(\boldsymbol{r},\boldsymbol{k}',t)] - S(\boldsymbol{k}',\boldsymbol{k})f(\boldsymbol{r},\boldsymbol{k}',t)[1-f(\boldsymbol{r},\boldsymbol{k},t)]\}, \qquad (2)$$

where $S(k, k')$ is the differential scattering rate from state $|k\rangle$ to state $|k'\rangle$. Vectors are denoted by bold characters. For a uniform electric field in a homogeneous system, the Boltzmann equation in the steady state reads:

$$\frac{1}{\hbar} F \frac{\partial f(k)}{\partial k} = I_c[f]$$

(3)

The equilibrium carrier distribution is simply given by the Fermi-Dirac occupation factor,

$$f_0(E) = \frac{1}{\exp\left[\frac{E-E_F}{k_B T}\right] + 1}$$

(4)

The Fermi level for an intrinsic system is found numerically by matching the concentrations of electrons and holes in the well, e.g., $\sum_i n_i = \sum_i 2/(2\pi)^2 \int f_0(E_i(k)) d^2\vec{k}$ and $\sum_j p_j = \sum_j 2/(2\pi)^2 \int (1 - f_0(E_j k)) d^2 k$. For a doped system with a given electron concentration, the Fermi level is found by fitting the electron concentration in the well $\sum_i n_i$ to the needed value. Here $E_i(k)$ and $E_j(k)$ are electron and hole dispersions calculated using the methods described in Section 1. For weak electric fields, we assume field-induced changes of the Fermi level to be negligible.

In the presence of an electric field, the distribution function $f$ undergoes an axially symmetric perturbation, with the axis determined by the field direction. In this case $f$ may be expanded in terms of the Legendre polynomials $P_n(cos\alpha)$, where $\alpha$ is the angle between $k$ and $F$ (for $k'$, this angle is denoted $\alpha'$) [10]:

$$f(k) = \sum_n f_n(E) P_n(cos\alpha)$$

(5)

For weak electric fields, it suffices to take into account only the first two terms of the series:

$$f(k) = f_0 - \frac{F}{\hbar} cos\alpha \frac{\partial E}{\partial k} \frac{\partial f_0}{\partial E} \phi(E).$$ (6)

Here $\phi(E)$ is the perturbation distribution, which has the physical dimension of time. One should note that the coefficients for the disturbed part of $f(k)$ can be written in arbitrary form because the actual dependence $\phi(E)$ remains to be found. The form (6) was chosen to simplify the further algebra. To obtain a linearized form of the collision integral (2), $f(k)$ in the form of (6) is substituted into the expression (2).

While simplifying (2), one needs several additional equations and assumptions. First, the principle of detailed balance was used: $S(k, k')f_0(E)[1 - f_0(E')] = S(k', k)f_0(E')[1 - f_0(E)]$. Second, due to the assumption of low electric field, only the terms linear in $F$ were retained. Third, $\alpha' = \alpha + \vartheta$ (where $\vartheta$ is the angle between k and $k'$), thus $cos\alpha' = cos\vartheta \cdot cos\alpha - sin\vartheta \cdot sin\alpha$. Fourth, $S(k, k')$ is symmetric with respect to the inversion of the sign of $\alpha'$. Consequently, terms with $\sin\vartheta$ vanish in the integral (2). Finally, a simplified form for the collision integral was derived:

$$I_c[f] = \frac{F}{\hbar}\frac{\partial f_0}{\partial E}\cos\alpha \times \int \frac{d^2\mathbf{k}'}{(2\pi)^2}\frac{1-f_0(E')}{1-f_0(E)} \times \left[\frac{\partial E}{\partial k}\phi(E) - \cos\vartheta\frac{\partial E'}{\partial k'}\phi(E')\right]S(\mathbf{k},\mathbf{k}')$$
(7)

Substituting (6), (7) into equation (3) and neglecting the term proportional to $F^2$ in the left-hand side of (3), one can derive the linearized Boltzmann equation (LBTE):

$$1 = \int \frac{d^2\mathbf{k}'}{(2\pi)^2}\frac{1-f_0(E')}{1-f_0(E)} \times \left[\phi(E) - \cos\vartheta\frac{\partial E'/\partial k'}{\partial E/\partial k}\phi(E')\right]S(\mathbf{k},\mathbf{k}')$$
(8)

It is important to note that in the LBTE, the formalism of effective mass was not used. Such energy-dependent effective mass $m(E)$ can be introduced via the usual relation used for the conductivity modeling [11, 12]:

$$m(E) = \frac{\hbar^2 k}{\partial E/\partial k}.$$
(9)

In the limiting case of a parabolic dispersion, the effective mass (9) becomes energy-independent, and LBTE (8) reduces to the usual form (see Eq.(9) of [10]).

## 3.2. Differential scattering rates

Consider three scattering mechanisms that are most important for MCT: longitudinal optical (LO) phonon scattering, charged impurities (CI) scattering and electron-hole (EH) scattering.

LO phonon scattering is inelastic, longitudinal optical phonons can be treated as having the constant energy $\hbar w_0$ (in calculations we have used the value of phonon energy for HgTe, $\hbar w_0 = 17\ meV$ [11]). Charged impurities scattering and electron-hole scattering are elastic and do not change the electron energy. Differential scattering rates for different scattering mechanisms are additive, thus the total differential scattering rate $S(\mathbf{k},\mathbf{k}')$ can be expressed as:

$$S(\mathbf{k},\mathbf{k}') = S_{CI}(\mathbf{k},\mathbf{k}')\delta(E,E') + S_{EH}(\mathbf{k},\mathbf{k}')\delta(E,E') + S_{LO}^a(\mathbf{k},\mathbf{k}')\delta(E+\hbar w_0, E') \\ + S_{LO}^e(\mathbf{k},\mathbf{k}')\delta(E-\hbar w_0, E')\theta(E-\hbar w_0 - E_0)$$
(10)

where $S_{LO}^a(\mathbf{k},\mathbf{k}')$ and $S_{LO}^e(\mathbf{k},\mathbf{k}')$ are the differential scattering rates for the phonon absorption and emission, while $S_{CI}(\mathbf{k},\mathbf{k}')$ and $S_{EH}(\mathbf{k},\mathbf{k}')$ are the differential scattering rates for charged impurities and electron-hole scattering, respectively. $\theta(E - \hbar w_0 - E_0)$ is the unit step function, which ensures there is no scattering at the energies below the bottom of the band.

Substituting (10) into (8), one can obtain:

$$1 = \phi(E) \int \frac{d^2\mathbf{k}'}{(2\pi)^2} \left\{ \frac{[1-f_0(E+\hbar w_0)]}{[1-f_0(E)]} S_{LO}^a(\mathbf{k},\mathbf{k}')\delta(E+\hbar w_0, E') \right.$$
$$+ \frac{[1-f_0(E-\hbar w_0)]}{[1-f_0(E)]} S_{LO}^e(\mathbf{k},\mathbf{k}')\delta(E-\hbar w_0, E')\theta(E-\hbar w_0 - E_0)$$
$$\left. + (1-\cos\vartheta)(S_{CI}(\mathbf{k},\mathbf{k}') + S_{EH}(\mathbf{k},\mathbf{k}'))\delta(E, E') \right\}$$
$$- \phi(E+\hbar w_0) \int \frac{d^2\mathbf{k}'}{(2\pi)^2} \frac{[1-f_0(E+\hbar w_0)]}{[1-f_0(E)]}$$
$$\times \left[\cos\vartheta \frac{\partial E'/\partial k'}{\partial E/\partial k}\right] S_{LO}^a(\mathbf{k},\mathbf{k}')\delta(E+\hbar w_0, E')$$
$$- \phi(E-\hbar w_0) \int \frac{d^2\mathbf{k}'}{(2\pi)^2} \frac{[1-f_0(E-\hbar w_0)]}{[1-f_0(E)]}$$
$$\times \left[\cos\vartheta \frac{\partial E'/\partial k'}{\partial E/\partial k}\right] S_{LO}^e(\mathbf{k},\mathbf{k}')\delta(E-\hbar w_0, E')\theta(E-\hbar w_0 - E_0)$$

(11)

One should note that $\delta(E,E')$ (which is equal to $\delta(k,k')$, note that $k$ and $k'$ are scalars), $\delta(E+\hbar w_0, E')$ and $\delta(E-\hbar w_0, E')$ reflect energy conservation laws that restrict the range of possible values of $k'$ in (11). Thus two-dimensional integrals over the whole 2D $k$-plane reduce to one-dimensional integrals over the angle $\vartheta$.

### 3.3. Differential scattering rate for charged impurities scattering

To obtain charged impurities differential scattering rate, we adopted the approach of [13] for nonzero temperatures. In *n*-type HgCdTe solid solutions, ionized charged impurities have the charge *e*. We assumed that residual charged impurities are located in the well, while the influence of barrier impurities is negligible.

The scattering rate of ground level electrons having the energy *E* is given by the equation similar to Eq.(94) of Ref.[13]:

$$\frac{1}{\tau(E)} = \int \frac{d^2\mathbf{k}'}{(2\pi)^2}(1-\cos\vartheta)S_{CI}(\mathbf{k},\mathbf{k}')\delta(E,E')$$
$$= \frac{m(E)}{\pi\hbar^3} c_{imp} \int_0^\pi d\vartheta(1-\cos\vartheta) \left[\frac{2\pi e^2}{\kappa\left[2k\sin\frac{\vartheta}{2} \cdot \xi_{2DEG}\left(2k\sin\frac{\vartheta}{2}, E\right)\right]}\right]^2$$
$$\times \int_{-L/2}^{L/2} dz \cdot g_{imp}^2\left(2k\sin\frac{\vartheta}{2}, z\right)$$

(12),

where $c_{imp}$ is the volume concentration of residual charged impurities, and the effective mass *m(E)* is defined by (9). The form-factor $g_{imp}$ is defined by the following formula (see p.213 of [13]):

$$g_{imp}(q_\perp, z) = \int_{-\infty}^{\infty} \chi^2(zv) \cdot \exp[-q_\perp |zv - z|] dzv ,$$

where $\chi^2(z)$ is the square of envelope function of the ground level electron, which is defined below. In (12), $\xi_{2DEG}$ is the screening function. To choose the proper screening function, one has to take into account that both electrons and holes take part in the screening. Heavy holes move much slower than light electrons, but in the zero frequency limit they nevertheless influence the screening of carriers in 2DEG. For that reason, standard formulas for the screening function that take into account only one type of carriers (see, e.g., Eqs.(68-71) of Ref. [13]), are poorly applicable to HgCdTe semi-metal systems. Further, in the doped case the Fermi level is in the conduction band, so the dispersion law at the Fermi level is almost linear, which should change the form of the screening function in comparison to the standard case with a parabolic dispersion law. For bulk HgTe, the appropriate screening function is given in [14], but it is unsuitable for our calculations since we deal with a two-dimensional electron gas. We used the screening function obtained by Hwand and Das Sarma for graphene [15]. Graphene is also a two-dimensional system with a linear dispersion law, and recent experiments [16] confirm the graphene-like screening properties of HgTe two-dimensional systems. This screening function has the following form [15]:

$$\xi_{2DEG}(q_\perp, E) = \begin{cases} 1 + \dfrac{2\pi e^2}{\varepsilon_0 q_\perp} DOS(k_F), & q_\perp < 2k_F \\ 1 + \dfrac{2\pi e^2}{\varepsilon_0 q_\perp} DOS(k_F)\left(1 + \dfrac{\pi q_\perp}{8k_F} - \dfrac{1}{2}\sqrt{1 - \dfrac{4k_F^2}{q_\perp^2}} - \dfrac{q_\perp}{4k_F}ArcSin\dfrac{2k_F}{q_\perp}\right), & q_\perp \geq 2k_F, \end{cases}$$
(13)

where $k_F$ is the modulus of the wave vector of the ground level electron at the Fermi level, $\varepsilon_0$ is the static permittivity of the lattice, and $DOS_{Ef}$ is the electron density of states at the Fermi level, $DOS(k) = \dfrac{1}{\pi}\dfrac{k}{\partial E/\partial k}$.

### 3.4. Differential scattering rate for electron-hole scattering

According to [14], the electron-hole scattering rate can be calculated similarly to the rate of electron scattering on charged impurities and is given by Eq. (12) where the concentration of charged impurities $c_{imp}$ should be replaced by the effective number of holes. Assuming that the value of $q_\perp$ is small compared to the average wave vector of holes, the effective number of holes can be written as [14]:

$$N_h(q_\perp) = \frac{1}{S}\sum_{k_h,n} f_0^h(k_h, n)(1 - f_0^h(|k_h - q_\perp|, n))$$
(14)

Growth of the Fermi energy enhances the screening (13), which partially suppresses this type of scattering. In contrast to the charged impurity scattering, for the electron-hole scattering both the screening function and the effective number of holes $N_h(q_\perp)$ strongly depend on the position of the Fermi level.

If the Fermi level is at least several $k_B T$ higher than the top of the hole band, heavy holes can be treated as non-degenerate. In this case, to a very good approximation, $(1 - f_0^h(|k_h - q_\perp|, n)) \approx 1$, and $N_h(q_\perp)$ is approximately equal to the concentration of holes. A further growth of the Fermi level results in a fast decrease of the EH scattering rate, because the concentration of holes drops as the screening function (13) grows.

On the other hand, if the Fermi level is situated lower than the top of the ground hole band, holes become degenerate. In this case, holes are scattering into states with a high occupancy factor, so $N_h(q_\perp)$ is much smaller than the concentration of holes in the quantum well. Consequently, the EH scattering will be strongly suppressed in this case as well.

These considerations on the EH scattering will be used in the next Section to explain the numerical results and the existing experimental data.

### 3.5. Differential scattering rate for longitudinal optical phonon scattering

For the LO phonon scattering, the relaxation rate of the carriers from the state $(k_0)$ is given by Eq. (6.141) of Ref. [17]. The total scattering rate can be expressed through differential scattering rates with the help of the following standard formula (see Eq.(6.136) of Ref. [17]): $1/\tau_s = \sum_k S_{LO}^a(\boldsymbol{k}, \boldsymbol{k}') + S_{LO}^e(\boldsymbol{k}, \boldsymbol{k}')$. Combining those formulas, one can see that the differential scattering rates $S^{e,a}(\boldsymbol{k}, \boldsymbol{k}')$ of the LO phonon absorption and emission can be written as:

$$S_{LO}^{e,a}(\boldsymbol{k}, \boldsymbol{k}') = \frac{(2\pi)^2}{2} \frac{2e^2 w_0}{L} \left( \frac{1}{\varepsilon_\infty} - \frac{1}{\varepsilon_0} \right) \left( N_{LO}(T) + \frac{1}{2} \pm \frac{1}{2} \right)$$
$$\cdot \sum_m \left\{ G(m) \frac{\delta[E(\boldsymbol{k}) - E(\boldsymbol{k}') \mp \hbar w_0]}{((\boldsymbol{k}' - \boldsymbol{k})^2 + (\pi m/L)^2)} \right\},$$
(15)

where $L$ is the QW width, $m$ is the phonon mode quantum number, $\varepsilon_0$ and $\varepsilon_\infty$ are static and high-frequency permittivities of the QW material, and $N_{LO}(T) = 1/(\exp[\hbar w_0/k_B T] - 1)$ is the phonon occupation number.

The form-factor $G(m)$ is given by Eq. (6.141) of Ref. [17]:

$$G(m) = \left( \int_{-L/2}^{L/2} \chi^2(z) \cos\left(\frac{\pi m z}{L}\right) dz \right)^2 \text{ for odd } m,$$
$$G(m) = \left( \int_{-L/2}^{L/2} \chi^2(z) \sin\left(\frac{\pi m z}{L}\right) dz \right)^2 \text{ for even } m.$$

Polar optical phonon scattering dominates over non-polar optical phonon scattering in HgTe [6, 14], so one has to choose a proper screening function. Following the approach for the screening function from Section 3.4, we use the results of Hwang and Das Sarma [15], obtained for the case of a linear energy dispersion law. In the case of phonons, static screening function becomes inapplicable and we use the dynamic screening function for the frequency of phonon. To include screening effects into the equation for the LO phonon differential scattering rate (15), this rate should be simply divided by the square of the screening function.

### 3.6. Evaluation of form-factors for the mixed bands

In narrow-gap degenerate semiconductor with a nonparabolic dispersion law, the electron wave function is a strong mix of states belonging to different bands. In the framework of the 8-band k.p method [4,5], the carrier wave function is expanded in the basis of eight Bloch band-edge (in-plane $k=0$) functions $u_n(r)$:

$$\psi(r) = exp[i(k_x x + k_y y)] \sum_{n=1}^{8} \chi_n(z) u_n(r),$$

where envelope functions $\chi_n(z)$ are found from the calculations of the energy spectrum [4]. According to the definition, $\chi_n(z)$ are slowly varied functions on the scale of the unit cell, while the Bloch functions vary rapidly on the scale of the unit cell. The value of $\chi^2(z)$ can be found using the fact that the Bloch functions form the full orthonormal basis:

$$\chi^2(z) = \int \psi(r) \cdot \psi^*(r) dr$$
$$= \int \left[\sum_{n=1}^{8} \chi_n(z) u_n(r)\right] \cdot \left[\sum_{n=1}^{8} \chi_n^*(z) u_n^*(r)\right] dz$$
$$\approx \int \sum_{n=1}^{8} \chi_n(z) \chi_n^*(z) \, dz$$

(16)

Calculation of the electron wave functions takes a lot of a computational time. For the sake of for simplicity, in our calculations of scattering rates in (12) and (15) we used wave functions, calculated at the Fermi level.

### 3.7. Iterative procedure for calculating the perturbation distribution $\phi(E)$

The LO phonon scattering is strongly inelastic because the LO phonon energy (17 meV) is several times larger than $k_B T$ (6.7 meV). As a result, a simple relaxation time approximation becomes inapplicable. In our calculations of the mobility, the linearized Boltzmann transport equation (LBTE) was solved directly by means of the iterative technique, which allows one to calculate perturbation distribution $\phi(E)$ for inelastic scattering.

The iterative procedure of calculating $\phi(E)$ is based on Eq. (11) and is described in [18]. In the first step of this iterative procedure one assumes that the "upper" term $\phi(E + \hbar w_0)$ and the "lower" term $\phi(E - \hbar w_0)$ are both equal to zero, then one can find $\phi(E)$ from the LBTE in a simple algebraic way. In the following steps, one uses the lower and upper terms from the previous iteration, $\phi_{n-1}(E - \hbar w_0)$ and $\phi_{n-1}(E + \hbar w_0)$ to find $\phi_n(E)$. The procedure continues until the difference between $\phi_{n-1}(E)$ and $\phi_n(E)$ falls within the needed tolerance bounds.

Despite its simplicity, this procedure demands a huge amount of computation time since its convergence is very slow. Therefore, we have used a modified version of the iterative procedure with a faster convergence. In this modification, the starting values for the procedure consisting of *(n+1)* iterations were determined as follows: $\phi_0(E - n\hbar w_0)$ was found from (11) as in the standard iteration procedure [18] (setting lower and upper terms to be zero); and $\phi_0(E + l\hbar w_0)$ (for *l>-n*) was found from (11) using $\phi_0(E + (l-1)\hbar w_0)$ as the lower term and taking the upper term to be zero. To find n-th order term $\phi_n(E)$ from (11) one uses $\phi_n(E - \hbar w_0)$ as the lower term (if it is already found, otherwise $\phi_{n-1}(E - \hbar w_0)$ is used), and $\phi_{n-1}(E + \hbar w_0)$ as the upper term.

In our calculations, the iterative procedure was repeated until the difference between $\phi_{n-1}(E)$ and $\phi_n(E)$ became smaller than 5%. The convergence was usually reached after 3-4 iterations.

## 4. Estimates of the effect of minor scattering mechanisms

### 4.1 Scattering on acoustic phonons

In HgCdTe quantum wells, there might be two channels of electron relaxation via the acoustic phonons, namely, the scattering on the deformation potential and the scattering by the piezoelectric interaction [17]. However, for the quantum wells with the (001) growth direction, the piezoelectric interaction is absent (see p.48 of Ref. [19]). Thus, in this Section we estimate the relaxation time of electrons which scatter on acoustic phonons via the deformation potential interaction only.

In layered heterostructures, acoustic waves consist of extended and confined modes. Extended modes propagate through the whole heterostructure in any direction. Confined modes are localized in the layer plane of the well and propagate along this plane. However, due to the small difference between the elastic properties of the matrix and the quantum well materials, such a confinement is weak. For that reason, the confined modes can be neglected, while extended modes can be approximated by plane waves. For HgTe quantum well of 20 nm width, the energy of acoustic phonon can be estimated as $E_{acoustic} \sim \hbar s_L(2\pi/L)$ (see Eq. (6.135) of Ref. [17]), where $s_L \approx 10^5$ cm/s is the sound velocity in the layer plane. The acoustic phonon energy is thus merely about 0.2 meV, so the scattering can be assumed to be elastic.

For our estimates we have used the simple model (see Eq. (6.137) of Ref. [17]), which assumes the well to be infinitely deep, electrons to scatter within the ground level only, acoustic phonons to be bulk-like, and neglects the nonparabolicity of the energy dispersion. This yields:

$$(\tau_s(k_F))^{-1} = \frac{2(C-a)^2 m(E_F) k_B T}{\hbar^3 s_L C_L L} \frac{3}{4} ,$$

$$\tag{17}$$

where the deformation potential constant $(C-a)_{HgTe} = -3.69\ eV$ [20], the mean elastic constant of the material $C_L = M_0 s_L/V_0 = 5.0625*10^{17}$ eV*s/cm$^4$ , and $M_0$ and $V_0$ are the mass and the volume of the unit cell of HgTe, respectively.

According to our estimates, for the intrinsic case the acoustic phonon relaxation time of electrons at the Fermi level is about $2*10^{-11}$ s for a well with the width $L = 20$ nm and $x = 0$, while for the well with $L = 20$ nm and $x = 0.09$ this time is around $6*10^{-11}$ s. These times are several orders of magnitude greater than the appropriate values of $\phi(E)$ at the Fermi energy. Thus, scattering on acoustic phonons is relatively small and can be neglected in the further treatment.

### 4.2 Scattering on the interface roughness

Analysis of [21] reveals that for semiconductors with a direct band order the mobility limited by the interface roughness (IR) scattering scales as $\mu_{IR} \sim L^6$ with the well width $L$. For semimetals with an inverted band order, the mobility determined by this type of scattering increases by about an order of magnitude (see the discussion in Ref. [21]).

Ref. [21] presents IR-scattering-limited mobilities in the semiconductor type HgTe/CdTe superlattices (with the QW width lower than the critical one). Semimetal-type $L=12$ nm HgTe and $L=20$ nm Hg$_{1-x}$Cd$_x$Te with the composition $x \leq 0.09$ (see Fig. 4) quantum wells from our work have inverted bands order according to our calculations. Extrapolating the data of Ref. [21] to greater QW widths according to the rule $\mu_{IR} \sim L^6$ and accounting the peculiarities of IR scattering mobility in inverted bands system outlined above, one can roughly estimate the IR-scattering-limited mobilities for Hg$_{1-x}$Cd$_x$Te quantum wells in our case. For $L=12$ nm QW, the estimated mobility is about $3*10^7$ cm$^2$/(Vs), while for $L=20$ nm QW, this mobility should be about $5*10^8$ cm$^2$/(Vs). Those numbers are much higher than the best mobility values (see Fig.6-8) given by other scattering mechanisms in our estimates. We conclude that the interface roughness scattering could be neglected in the mobility calculations.

### 4.3. Scattering on fluctuations of composition and effective mass

For wide QWs, fluctuations of the well width are small compared to the well width itself (monolayer fluctuations at the interface have the amplitude of approximately 0.5 nm). Normally, the corresponding fluctuations of the effective band gap and effective mass are also small in relative units; however, for the critical well width, even a small fluctuation of the QW width could cause a significant change of the electron effective mass. It is also important to note, that Hg$_{1-x}$Cd$_x$Te heterostructures are grown with the precision of composition $\Delta x = \pm\ 0.002$ [22]. At the critical well width, these composition fluctuations also can cause considerable changes of the electron effective mass. Thus, near the band inversion point, the approach outlined in the present article should be revised to include the effect of additional types of scattering.

## 4.4. Scattering on the alloy disorder

The alloy disorder scattering originates from disturbances of the periodic HgTe lattice introduced by Cd atoms. This scattering is completely absent for pure HgTe, and it is more pronounced for samples with higher composition $x$.

In our case, the maximum alloy disorder scattering (and the minimum of the relevant mobility) occurs for the maximal considered value $x = 0.09$. We estimated the impact of such a scattering mechanism on the electron mobility for $L=20$ nm QW with $x=0.09$, using the methods of Ref. [13] (see p. 216). As the alloy disorder scattering is screened (see, for example, Refs. [23, 24]), the estimated values of this mobility are about $10^8$ cm$^2$/(Vs) in the entire range of electron concentrations. These estimates are in a good agreement with the results for bulk MCT presented in Fig.4 of Ref. [6], where the alloy-disorder-limited mobility for much larger $x$ ($x=0.17$) is reported to be higher than $10^7$ cm$^2$/(Vs) at T=77K. Thus, we conclude that this scattering mechanism can be neglected as well.

## 5. Electron mobility – calculations and discussion

Under the action of an external dc electric field $\boldsymbol{E}_{\text{ext}}$, the distribution function is deformed from $f_0$ to $f(k)$ given by Eq. (6), and the average electron velocity $\langle v \rangle$ in the 2DEG becomes nonzero. This creates the electric current with the density $\boldsymbol{j} = en\langle \boldsymbol{v} \rangle$. Using the definition of the electron velocity $\boldsymbol{v} = \frac{1}{\hbar}\frac{\partial E}{\partial \boldsymbol{k}}$ [25], the current density $j$ is found by averaging all possible electron velocities in the QW:

$$\boldsymbol{j} = \frac{2}{(2\pi)^2}\frac{e}{\hbar}\int \frac{\partial E}{\partial \boldsymbol{k}} f(\boldsymbol{k}) d^2\boldsymbol{k}$$

(19)

The drift mobility can be found from its definition $\mu_D = j/(n_s F)$ (note that only the disturbed part of the distribution function (6) enters the integral below):

$$n_s = \frac{2}{(2\pi)^2}\int f(k) d^2\vec{k},$$

$$\mu_D = \pi \frac{1}{2\pi^2}\frac{e}{\hbar^2}\frac{1}{n_s}\int \left(\frac{\partial E}{\partial k}\right)^2 \frac{\partial f_0}{\partial E}\phi(E) \cdot k \cdot dk$$

(20)

To check the results obtained, first of all we compare them with the available experimental data from the literature. Since, to our knowledge, there is no detailed experimental data on the electron mobility in the system under consideration at $T =$ 77 K, we will compare our numerical results with the experiments done at $T = 4.2$ K [26] and discuss qualitative features.

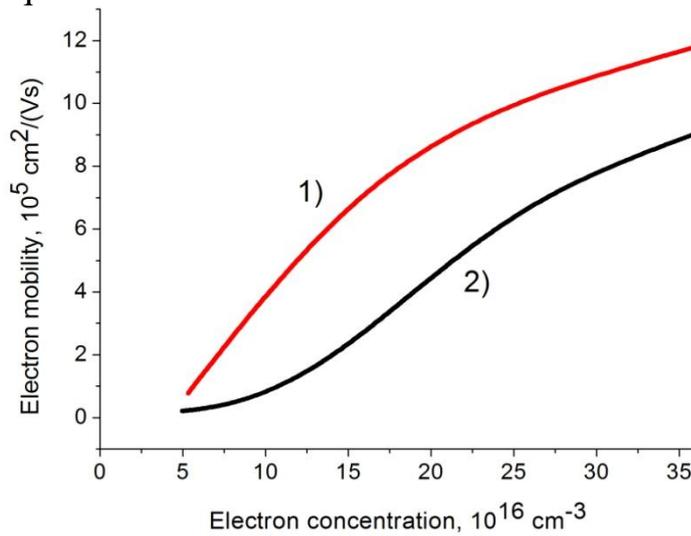

Figure 5. Dependence of the mobility on the electron concentration in the $Hg_{0.3}Cd_{0.7}Te/HgTe/Hg_{0.3}Cd_{0.7}Te$ QW with the width of 12 nm. Curve 1 denotes the experimental data from [26] for T = 4.2 K. Curve 2 shows our numerical data for T = 77 K. Concentration of residual charged impurities in the well is $1.14*10^{15}$ cm$^{-3}$ (corresponds to $1.09*10^{10}$ cm$^{-2}$).

Results of our numerical calculations for the electron mobility in a 12 nm wide HgTe quantum well at $T = 77$ K are presented in Fig.5 (curve 2). We compare those results with the experimental data of Tkachov et al [26] obtained at $T = 4.2$ K: curve 1 in our Fig.5 corresponds to the curve for sample #6 in Fig.2 of Ref. [26]. In Ref.[26], the electron concentration in the QW was changed by applying the top-gate bias. One can see that mobility dependencies, shown by the two curves in Fig.5, qualitatively agree with each other. Both the experimental and theoretical curves exhibit low mobility at low electron concentration; the monotonic growth of the mobility with the increase of the electron concentration can be explained by the decrease of the hole concentration and the enhancement of screening. However, the experimental mobility from [26] (measured at liquid helium temperature) increases faster with the electron concentration than our simulated mobility calculated at liquid nitrogen temperature. This probably could be explained by a faster decrease of the hole concentration with the Fermi level shift at $T = 4.2$ K.

Although a direct quantitative comparison of the mobilities obtained for different temperatures is not meaningful since most of the dominant scattering mechanisms strongly depend on temperature, there are experiments which show that in HgTe QWs the mobility at liquid helium temperature is 1.5-2 times larger than the mobility at liquid nitrogen temperature [27,28], which is in a qualitative agreement with our Fig.5.

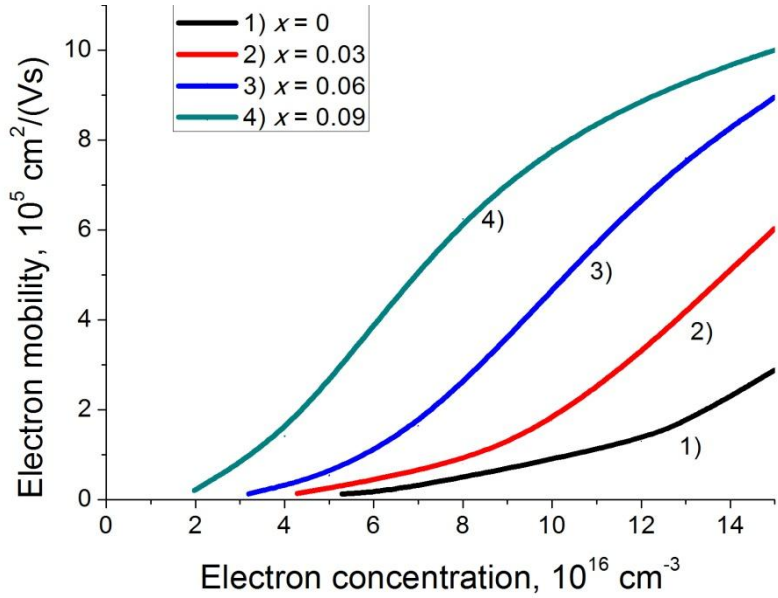

Figure 6. Dependence of the mobility on the electron concentration in $Hg_{0.32}Cd_{0.68}Te/Hg_{1-x}Cd_xTe/Hg_{0.32}Cd_{0.68}Te$ QW. Curves 1, 2, 3, 4 correspond to molar compositions $x = 0$, 0.03, 0.06 and 0.09, respectively. The QW width is 20 nm. The concentration of residual charged impurities in the well is $10^{15}$ cm$^{-3}$.

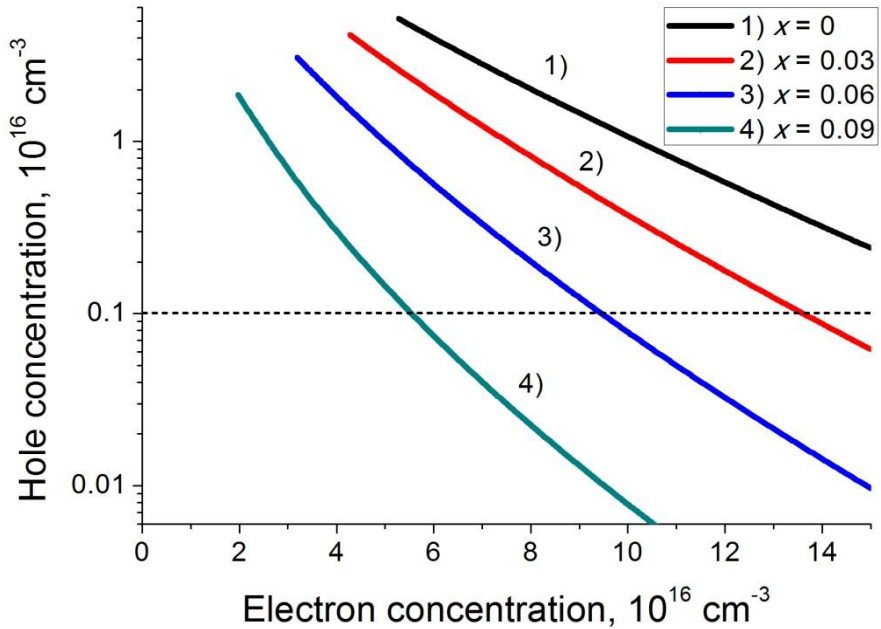

Figure 7. Dependence of the hole concentration on the electron concentration in $Hg_{0.32}Cd_{0.68}Te/Hg_{1-x}Cd_xTe/Hg_{0.32}Cd_{0.68}Te$ QW of 20 nm width. Curves 1, 2, 3, 4 correspond to molar compositions $x = 0$, 0.03, 0.06, 0.09, respectively. The leftmost point in each curve corresponds to the intrinsic case (cf. Fig. 6). The horizontal dotted line indicates the concentration of residual charged impurities in the QW.

Our numerical results for the electron mobility in a 20 nm wide $Hg_{0.32}Cd_{0.68}Te/Hg_{1-x}Cd_xTe/Hg_{0.32}Cd_{0.68}Te$ quantum well at $T = 77$ K, for different compositions $x$, are presented in Fig. 6, while Fig. 7 shows the corresponding dependences of the hole concentration on the electron concentration in the QW, which can be adjusted by barrier doping or by applying the top-gate bias.

Consider impacts of each scattering mechanism. Strong dynamical screening leads to a strong suppression of the longitudinal optical phonon scattering. For an intrinsic 20 nm wide quantum well with the composition $x$=0, the electron mobility for the LO phonon scattering is about $3.8*10^6$ cm$^2$/(Vs), while for n-doped quantum well of the same width with composition $x$=0.06 (electron concentration $1.5*10^{17}$ cm$^{-3}$) the mobility is about $6.8*10^6$ cm$^2$/(Vs). As these mobilities are much greater than the corresponding total mobilities (see Fig.6), we can conclude that the main contribution to the total mobility comes from the charged impurity scattering and electron-hole scattering. Relative importance of these two scattering mechanisms can be easily seen from the comparison of the hole and charged impurity concentrations presented in Fig.7.

From Fig.6 one can see that the increase of the electron concentration leads to the increase in the electron mobility. This could be explained by two simultaneous processes – decrease of heavy hole concentration (see Fig.7) and enhancement of screening. The first process suppresses the electron scattering on heavy holes, while the second one suppresses all three scattering mechanisms considered.

At high electron concentrations the mobility growth becomes slower; this effect is more pronounced for higher compositions $x$. This fact could have the following explanation: First, there exists a competing process, which decelerates the mobility growth, namely the increase of the electron effective mass as the Fermi level goes up. Second, the heavy hole concentration (as a function of the electron concentration) decreases faster for quantum wells with higher $x$ (see Fig.7). For example, at $n = 10^{17}$ cm$^{-3}$, for HgTe QW (Fig.6, curve 1) the concentration of heavy holes is equal to $1.07*10^{16}$ cm$^{-3}$, while for Hg$_{0.91}$Cd$_{0.09}$Te quantum well (Fig.6, curve 4) it is equal to $7.8*10^{13}$ cm$^{-3}$. Taking into account that there is a background of charged impurities with a constant concentration of $10^{15}$ cm$^{-3}$ (see the dotted line in Fig.7), one should conclude that in quantum wells with a higher molar composition $x$, the total number of charged centers stops to decrease earlier than in QWs with lower $x$ (see Fig.7). Then one of the mobility enhancement mechanisms becomes ineffective. Growth of the mobility with the increase of the quantum well composition $x$ could be explained by a lower concentration of heavy holes at the same value of the electron concentration. From the above arguments, it is clear that in many cases the concentration of holes in QW is greater than $10^{15}$ cm$^{-3}$, and hence the growth of high purity samples with low concentration of residual charged impurities (of the order of $10^{14}$ cm$^{-3}$) will not improve the electron mobility sufficiently.

The magnitude of the drop in the electron mobility at low electron concentrations (see Figs.5,6) depends on the band structure type of the quantum well. This assertion can be illustrated by the comparison with the results of [29]: In contrast to the band structure with a finite gap (Ref. [26] and the present work, for the samples grown in (001) crystallographic plane), in the band structure considered in Ref. [29] (the sample with (013) crystallographic plane) the heavy-hole band and the conduction band overlap (see Fig.1 of Ref. [29]). Then in the intrinsic case the Fermi level lies both in the conduction and the heavy-hole bands, and reducing the electron concentration leads not only to the increase of the hole concentration, but to the degeneration of the hole subsystem as well. As a result, for the structures with overlapping bands, the drop in the electron mobility with the decreasing electron concentration is not so pronounced (see Fig.6(b) of Ref. [29]).

A direct comparison of our results with the existing experimental data on the electron mobility at $T = 77$ K is problematic, because there is very few experimental data about the electron mobility in HgTe quantum wells at liquid nitrogen temperatures, and all of them, to our knowledge, lack the information on electron concentrations in the QW. Nevertheless, one can compare the order of magnitude of the mobility values, observed in several different experiments, with our theoretical prediction (Fig. 6, curve 1) where the mobility varies roughly from $10^4$ to $2.5*10^5$ cm$^2$/(Vs) for electron concentrations varying in the range $(5\div15)*10^{16}$ cm$^{-3}$.

The authors of Ref. [30] measured the electron mobility in *n*-type Hg$_{0.05}$Cd$_{0.95}$Te/HgTe/Hg$_{0.05}$Cd$_{0.95}$Te superlattice, grown on the (112) oriented Cd$_{0.96}$Zn$_{0.04}$Te substrate. Quantum well and barrier widths were 8 nm and 7.7 nm, respectively. The mobility at 77 K was $1.15*10^5$ cm$^2$/(Vs).

The authors of Ref. [27] measured the mobility temperature dependence in the sample of 124 layers, with the HgTe quantum well thickness $d_1$=8.6 nm and the CdTe barrier thickness $d_2$=3.2 nm. At $T = 4.2$ K, the Hall mobility was $2.5*10^5$ cm$^2$/(Vs). The temperature dependence of electron mobility showed that in the temperature range $T = 4.2 - 40$ K, the mobility changed weakly, while at $T = 77$ K it decreased to $1.4*10^5$ cm$^2$/(Vs).

The authors of Ref. [28] measured the electron mobility for *p*-type HgTe(7.8 nm)/CdTe(2.9 nm) superlattice and obtained the value $6.5*10^4$ cm$^2$/(Vs).

Comparing our results to those of [30, 27] and [28], one has to to exercise some caution since the carrier mobility could be eventually impaired by the interface scattering due to a very narrow width of the QWs used in those experiments, while in the present paper we are considering wider wells.

On the whole, one can conclude that our numerical results yield correct order of magnitude for the electron mobility and are reasonably consistent with the existing experimental data for HgTe QWs.

## 6. Application to hot-electron bolometers

We have applied the calculations of the previous section to model the resistivity of the Hg$_{1-x}$Cd$_x$Te QW used as a channel of the THz-range hot-electron bolometer. In this case, the channel thickness corresponds to the quantum well width $L$. Using the lateral dimensions of the bolometer channel (width $D_w$ and length $D_l$), one can calculate the channel resistivity $\rho$ in a usual way as $\rho = 1/(e\mu n)$ and its resistance is then given by $R = \rho D_l/(D_w L)$. Fig. 8 presents the results of such calculation for a channel with $L$=20nm and $D_w = D_l$=50 μm.

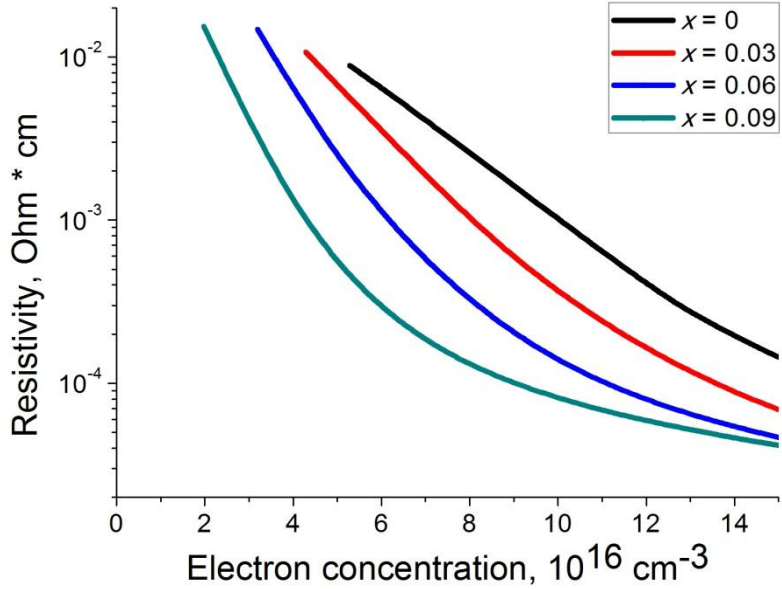

Figure 8. Dependence of the channel resistivity on the electron concentration, for different molar composition $x$ in $Hg_{0.32}Cd_{0.68}Te/Hg_{1-x}Cd_xTe/Hg_{0.32}Cd_{0.68}Te$ quantum well of 20 nm width. The concentration of background charged impurities in the channel is $10^{15}$ cm$^{-3}$.

From Fig.8, one can see that the channel resistivity varies by more than two orders of magnitude (from several tens of Ohm to about 10 kOhm) depending on the electron concentration. Such a strong dependence of the channel resistance on the electron concentration could provide high volt-watt sensitivity of the hot-electron bolometer, as small variations in the gate voltage should result in strong changes of the bolometer resistance. One more benefit of the considered system for the needs of THz detection lies in the high dynamical tunability. Bias voltage applied to the top-gate could change the channel resistance by several orders of magnitude, crucially affecting operational characteristics of the entire device. Such a tunable resistance makes it easy to match it electrically to the impedance of antenna, which increases the detector sensitivity.

There are three factors which determine the preferable value of the channel resistance for a THz hot-electron bolometer integrated with an antenna. First, a too high resistance causes a high thermal noise that leads to a lower NEP. Second, a resistance increase leads to a higher output signal at the same bias. Third, the channel resistance should be about the impedance of the integrated antenna (~ 100 Ohm) to provide the efficient energy transfer.

At frequencies over 0.5 THz, the efficiency of the incident radiation absorption by 2DEG decreases according to the Drude model. Thus, alternative ways should be used to improve the interaction of 2DEG with the incident THz radiation. One of the approaches to this problem is the use of plasma modes, which transfer the wave energy to 2DEG [31]; the other approach makes use of the higher QW doping to reduce the electron momentum relaxation time [2, 32]. The first approach needs high-mobility (undoped) channels, while the second approach needs lower-mobility channels with a fast momentum relaxation.

For the first approach [31] and frequencies about 1 THz, momentum relaxation times in the channel should satisfy the condition $\omega\tau > 1$, i.e., $\tau \sim 10^{-12} - 10^{-11}$ s. Considering the perturbation distribution $\phi(E)$ as a qualitative estimate for the relaxation time, we can conclude that this requirement could be fulfilled for high concentration of electrons in the channel. For such relaxation times, plasma waves will be excited. $\phi(E)$ can be considered as a qualitative estimate for the relaxation time, because according to the data of Section 5 inelastic phonon scattering is strongly suppressed, and the dominant scattering mechanisms are elastic.

For the second approach [2, 32] and frequencies about 1 THz, from the requirement $\omega\tau \ll 1$ it follows that an efficient interaction of 2DEG with the incident THz radiation could be achieved for $\tau \sim 10^{-13}$ s. Similar estimates lead us to the conclusion that this goal be reached for low composition $x$ and intrinsic carrier concentration in the channel.

Thus, in the case of high frequencies (about 1 THz), the preferred characteristics of the channel are determined by one of the approaches mentioned above. At lower frequencies (about 100 GHz), the preferable characteristics of semiconductor HEB channel could be considered from the standpoint of the electrical matching of the bolometer with the antenna.

## 7. Comparison of semi-metallic $Hg_{1-x}Cd_xTe$ quantum wells and graphene

In comparison to graphene, we can outline several benefits of using semi-metallic $Hg_{1-x}Cd_xTe$ quantum wells as a channel in THz HEB and FETs that could be of significant importance for the design of improved detectors.

First, graphene mobility significantly deteriorates when the graphene sheet is placed on a substrate. Its mobility can further decrease when a top-gate is grown on the graphene channel [33]. While high values ($10^6$ cm$^2$/(Vs)) of the mobility are measured at room temperature in exfoliated graphene sheets [34], substrate presence decreases the room-temperature mobility to measurable values of about (1...2.3)*$10^4$ cm$^2$/(V·s). Cooling down to the liquid nitrogen temperature increases the graphene mobility only slightly [35, 36] (particularly, see the inset in Fig. 3 of Ref. [35]). Thereby, devices built on $Hg_{1-x}Cd_xTe$ QWs can provide an order of magnitude higher mobility compared to graphene, which is of crucial importance, for example, for plasmonic applications.

Second, usually the graphene channels for HEBs are characterized by high resistances (typical resistances are of the order of 10 kOhm [37, 38]), which results in a bad matching with the antenna. In contrast to graphene, $Hg_{1-x}Cd_xTe$ QWs impedance can be matched to the antenna impedance, which should significantly increase the detector responsivity.

Third, due to the lower resistance, semi-metallic $Hg_{1-x}Cd_xTe$ channel will have lower thermal noise than a similar graphene channel. The thermal noise (the Johnson-Nyquist noise) is proportional to the square root of the resistance and is not negligible in semiconductor HEBs [39], so in a $Hg_{1-x}Cd_xTe$ channel one can reach several times lower levels of the thermal noise than in a graphene channel.

Fourth, one of the drawbacks of graphene as a channel for THz HEB lies in its very inefficient 2DEG excess energy relaxation. The dominant inelastic scattering mechanism that can provide energy relaxation is the optical phonon scattering [19], but the energy of an optical phonon in pure graphene is of the order of 200 meV [40]. Under THz radiation, this results in a decrease in the efficiency of the 2DEG energy relaxation, which could deteriorate the detector operation speed and degrade the overall performance of the detector. Comparing to graphene, 2DEG energy relaxation in semi-metallic $Hg_{1-x}Cd_xTe$ QW is much faster due to low energy of the LO phonon (17 meV in HgTe), since this energy and mean electron energy are of one order. Fast 2DEG energy relaxation could be important for increasing the detector operation speed.

## 8. Conclusions

We have studied the dependence of electron mobility in *n*-type semi-metallic $Hg_{1-x}Cd_xTe$ quantum wells on the electron concentration. In our simulations, processes of electron scattering on longitudinal optical phonons, charged impurities, and holes have been included. Principal features like band mixing, nonparabolicity of the dispersion law, and inelasticity of LO phonon scattering have been taken into account in the iterative solution of the linearized Boltzmann transport equation. We have also estimated the contributions from other scattering mechanisms involving acoustic phonons, interface roughness, alloy disorder, fluctuations of composition and effective mass, which have been found to be negligible for QW widths larger than 12 nm.

Comparing the separate impacts of each scattering mechanism, one can see that the longitudinal optical phonon scattering is strongly suppressed because of the strong dynamical screening. For an intrinsic 20 nm wide quantum well with the composition *x*=0, the electron mobility for the LO phonon scattering is about $3.8*10^6$ cm$^2$/(Vs), while for n-doped quantum well of the same geometry with composition *x*=0.06 (the electron concentration $1.5*10^{17}$ cm$^{-3}$), the electron mobility limited by the LO phonon scattering is about $6.8*10^6$ cm$^2$/(Vs). As these mobilities are much higher than the corresponding total mobilities, we can conclude that the main contribution to the total mobility comes from the charged impurity scattering and electron-hole scattering. Relative importance of these two scattering mechanisms can be established from the comparison of hole and charged impurity concentrations.

Our modeling has shown that at the liquid nitrogen temperature a high electron mobility can be obtained at high electron concentration in the well, which enhances 2DEG screening and decreases holes concentration. Such an increase of the electron concentration could be achieved by delta-doping of barriers or by applying the top-gate bias voltage.

Growth of the mobility with the increase of the quantum well composition *x* could be explained by a lower concentration of heavy holes at the same value of the electron concentration. Since the concentration of holes in QW is often higher than $10^{15}$ cm$^{-3}$, the fabrication of high purity samples with low concentration of residual charged impurities (of the order of $10^{14}$ cm$^{-3}$) will not improve the electron mobility sufficiently. Our modeling shows that because of the high hole concentration, the purity of samples in many configurations is of a lower importance for obtaining high electron mobility than the electron concentration in the well. This conclusion could be important for the reduction of fabrication costs for high-mobility HgCdTe heterostructures.

Our estimate of the resistance in semi-metal HgCdTe quantum wells used as a channel of the THz hot-electron bolometer shows that the channel resistance varies by more than two orders of magnitude depending on the electron concentration. Such a strong dependence could provide high volt-watt sensitivity of the hot-electron bolometer, as small variations in the gate voltage should result in strong changes of the bolometer resistance. A high dynamical tunability makes up another benefit of the considered system for the THz detection.

We have also assessed advantages of the HgCdTe THz hot-electron bolometer compared to the graphene HEB. We conclude that HgCdTe semi-metallic QWs can demonstrate higher mobility, lower thermal noise, higher operational speed, and can provide much more efficient coupling to planar antennas in THz range detector applications.

## 9. References

1. A. Rogalski, and F. Sizov, *Optoelectron. Rev.* **19**, 346 (2011).

2. J. K. Choi, V. Mitin, R. Ramaswamy, V.A. Pogrebnyak, M.P. Pakmehr, A. Muravjov, M.S. Shur, J. Gill, I. Mehdi, B.S. Karasik, and A.V. Sergeev, *IEEE Sensors Journal* **13**, 80 (2013).

3. E. B. Olshanetsky, Z. D. Kvon, Ya. A. Gerasimenko, V. A. Prudkoglyad, V. M. Pudalov, N. N. Mikhailov, and S. A. Dvoretsky, *JETP Letters* **98**, 843 (2013).

4. E.O. Melezhik, J.V. Gumenjuk-Sichevska, S.A. Dvoretskii, *Semiconductor Physics, Quantum Electronics & Optoelectronics* **17**, 179 (2014).

5. E.G. Novik, A. Pfeuffer-Jeschke, T. Jungwirth, V. Latussek, C.R. Becker, G. Landwehr, H. Buhmann, and L.W. Molenkamp, *Phys. Rev. B* **72**, 035321 (2005).

6. J.J. Dubowski, T. Dietl, W. Szymanska, R.R. Galazka, *J. Phys. Chem. Solids* **42**, 351 (1981).

7. M. König, S. Wiedmann, C. Brüne, A. Roth, H. Buhmann, L.W. Molenkamp, X.-L. Qi, Sh.-Ch. Zhang, *Science* **318**, 766 (2007).

8. B.A. Bernevig, T.L. Hughes, Sh.-Ch. Zhang, *Science* **314**, 1757 (2006).

9. V. Latussek, Ph.D. thesis, Julius-Maximilians-Universität, Würzburg, 2004.

10. T. Kawamura and S. Das Sarma, *Phys. Rev. B* **45**, 3612 (1992).

11. A.V.Liubchenko, E.A. Salkov, F.F. Sizov, Physical Bases of Semiconductor Infrared Photoelectronics (NaukovaDumka, Kiev, 1984) (In Russian).

12. C.Jacoboni, Theory of Electron Transport in Semiconductors (Springer, Berlin-Heidelberg, 2010) p. 120.

13. G. Bastard, Wave Mechanics Applied to Semiconductor Heterostructures (Halsted Press, New York, 1988).

14. W. Szymanska and T. Dietl, *J. Phys. Chem. Solids* **39**, 1025 (1978).

15. E. H. Hwang and S. Das Sarma, *Phys. Rev. B* **75**, 205418 (2007).

16. C.Brüne, C.Thienel, M.Stuiber, J.Böttcher, H. Buhmann, E. G. Novik, Ch.-X. Liu, E. M. Hankiewicz, and L. W. Molenkamp, Phys. Rev. X **4**, 041045 (2014).

17. V. Mitin, A. Kochelap, A. Stroscio, Quantum Heterostructures: microelectronics and optoelectronics (Cambridge University Press, Cambridge, 1999).

18. P. K. Basu and B.R. Nag, *Phys. Rev B* **22**, 4849 (1980).

19. M.A. Kinch, *Fundamentals of Infrared Detector Materials*, SPIE Press, Bellingham, Washington, 2007.

20. V. Latussek, C.R. Becker, G. Landwehr, R. Bini and L. Ulivi, *Phys. Rev. B* **71**, 125305 (2005).

21. J. R. Meyer, D. J. Arnold, C. A. Hoffman, and F. J. Bartoli, Appl. Phys. Lett. 58 (22), 2523 (1991)

22. S.A. Dvoretsky, D.G. Ikusov, Z.D. Kvon, N.N. Mikhailov, V.G. Remesnik, R.N. Smirnov, Yu.G. Sidorov, V.A. Shvets, *Semiconductor Physics, Quantum Electronics &Optoelectronics* **10**, 47 (2007).

23. A. Gold, JETP Letters 98(7), 416–420 (2013)

24. U. Bockelmann, G. Abstreiter, G. Weimann, W. Schlapp, PRB 41(11), 7864 (1990).

25. J. Singh, *Modern Physics for Engineers,* WILEY-VCH, Weinheim, 2004, Appendix B.

26. G. Tkachov, C. Thienel, V. Pinneker, B. B¨uttner, C. Br¨une, H. Buhmann, L. W. Molenkamp, and E. M. Hankiewicz, *Phys. Rev. Lett.* **106**, 076802 (2011).

27. L. Pyshkin and J. M. Ballato (Eds.), *Optoelectronics - Advanced Materials and Devices,* InTech, Rijeka, Croatia, 2013 Ch.6, p. 137.

28. J.R. Meyer, D.J. Arnold, C.A. HofFman, and F.J. Bartoli, *Phys. Rev. B* **46**, 4139 (1992).

29. M.V. Entin, L.I. Magarill, E.B. Olshanetsky, Z.D. Kvon, N.N. Mikhailov, S.A. Dvoretsky, Journal of Experimental and Theoretical Physics **117**, 933 (2013).

30. Y.D. Zhou, C.R. Becker, Y. Selamet, Y. Chang, R. Ashokan, R.. Boreiko, T. Aoki, D.J. Smith, A.L. Betz, and S. Sivananthan, *Journal of Electronic Materials* **32**, 608 (2003).

31. W. Knap, M. Dyakonov, D. Coquillat, F. Teppe, N. Dyakonova, J. Łusakowski, K. Karpierz, M. Sakowicz, G. Valusis, D. Seliuta, I. Kasalynas, A. El Fatimy, Y.Meziani, T. Otsuji, *Journal of Infrared, Millimeter, and Terahertz Waves* **30**, 1319 (2009).

32. R. Ramaswamy, K. Wang, A. Stier, A. Muraviev, G. Strasser, A. Markelz, M. Shur, R. Gaska, A.Sergeev, and V. Mitin, *Proc. SPIE, Micro- and Nanotechnology Sensors, Systems, and Applications III*, **8031**, 80310H (2011).

33. S. Kim, J. Nah, I. Jo, D. Shahrjerdi, L. Colombo, Zh. Yao, E.Tutuc, and S. K. Banerjee, *Applied Physics Letters* **94**, 062107 (2009).

34. F.Schwierz, *Nature Nanotechnology* **5**, 487 (2010).

35. B. Fallahazad, S. Kim, L. Colombo and E. Tutuc, *Appl. Phys. Lett.* **97**, 123105 (2010).

36. S.F. Chowdhury, S. Sonde, S. Rahimi, Li Tao, S. Banerjee and D. Akinwande, *Appl. Phys. Lett.* **105**, 033117 (2014).

37. Xu Du, D.E. Prober, H. Vora, C.B. Mckitterick, *Graphene and 2D Materials* **1**, 1 (2014).

38. D.B. Farmer, Hsin-Ying Chiu, Yu-Ming Lin, K.A. Jenkins, Fengnian Xia and P. Avouris, *Nano Lett.* **9**, 4474 (2009).

39. V. Dobrovolsky, F. Sizov, V. Zabudsky, N. Momot, *Terahertz Science and Technology* **3**, 33 (2010).

40. S. Winnerl, M. Orlita, P. Plochocka, P. Kossacki, M. Potemski, T. Winzer, E. Malic, A. Knorr, M. Sprinkle, C. Berger, W.A. de Heer, H. Schneider and M. Helm, *Phys. Rev. Lett.* **107**, 237401 (2011).